\documentclass[11pt,draftcls,onecolumn]{IEEEtran}   

\usepackage{amssymb}
\usepackage{amsmath}
\usepackage{amsfonts}
\usepackage{cases}
\usepackage{multirow}
\usepackage{graphicx}
\usepackage{caption}
\usepackage{array}
\usepackage{algorithm}
\usepackage{algorithmic}

\ifCLASSINFOpdf
\else
\fi
\begin{document}
%
\title{A Fast HRRP Synthesis Algorithm with Sensing Dictionary in GTD Model}
%
%

\author{R. Fan, Q. Wan, X. Zhang, H. Chen, Y.P. Liu \\uestc\_fanrong@foxmail.com, wanqun@uestc.edu.cn, z-xiao11@mails.tsinghua.edu.cn, huichen\_uestc@yahoo.cn, yipeng.liu@esat.kuleuven.be}


%
%

\markboth{Draft}%
{Shell \MakeLowercase{\textit{et al.}}:A Fast HRRP Synthesis Algorithm with Sensing Dictionary in GTD Model}
%



\maketitle


\textbf{Abstract---}
To achieve high range resolution profile (HRRP), the geometric theory of diffraction (GTD) parametric model is widely used in stepped-frequency radar system. In the paper, a fast synthetic range profile algorithm called orthogonal matching pursuit with sensing dictionary (OMP-SD), is proposed. It formulates the traditional HRRP synthetic to be a sparse approximation problem over redundant dictionary. As it employs a priori information that targets are sparsely distributed in the range space, the synthetic range profile (SRP) can be accomplished even in presence of data lost. Besides, the computational complexity is reduced by introducing sensing dictionary (SD) and it mitigates the model mismatch at the same time. The computation complexity decreases from $O(MNDK)$ flops for OMP to $O(M(N+D)K)$ flops for OMP-SD. Simulation experiments illustrate its advantages both in additive white Gaussian noise (AWGN) and noiseless situation, respectively.

%
\IEEEpeerreviewmaketitle

%
%
%
%
\textbf{1. ~Introduction}\par
\IEEEPARstart{A}{~HRRP} is the phasor sum of  the time returns from different scatterers on the target located within a resolution cell. From a geometric point of view, a HRRP represents the projection (in range) of  the apparent scattering centers onto the  radar  line  of  sight (LoS). It has been used into radar signal processing \cite{1}-\cite{5}. In GDT model, the radar target is no longer a point but composed of multiple scatterers along with radar LoS. Different targets are characterized by different scattering mechanisms and they are expressed as a function of frequency. From the perspective of digital signal processing, a process to identify radar target in GTD model is just the same as the process to estimate model parameters (containing scattering mechanisms, intensity and scatter range cells). In other words, it is also a process to estimate parameters of position and magnitude of target scatterers in range gate. While, in realistic process of synthesizing range profile, the returned signals are always inevitablely interfered by passive or/and active jamming. For the case, many returned signals are corrupted or even invalid. If signal sample is invalid, it has to be discarded. If exploiting the measured echo signals to synthesize range profile with traditional SRP algorithms directly, such as \cite{6}-\cite{7}, the SRP of target is either incorrect or flashed from coherence processing internal (CPI) to another one. Although interpolation or extrapolation strategy are helpful to improve the SRP, the range resolution remains constrained by system band width. However, it's worth noting that the significant physical scatterers are sparse in actual targets, which implies that strong scattering cells are also sparse for the target's SRP \cite{8}. This is consistent with sparse signal representation of compressed sensing (CS) theory appeared in recent years \cite{9}-\cite{13}. Hence, in the signal representation point of view, synthesizing range profile is equivalent to recovering a high-dimensional sparse signal form a low-dimensional measurements, usually accompanied by samples loss. It is an undetermined system. This will be confirmed in next sections.\par
In the past few years, some algorithms have been developed to solve an undetermined system by using sparse property. They are generally grouped into two categories in CS community. i) minimum $l_1$ norm reconstruction, i.e., optimization based on the $l_1$ norm can exactly recover sparse signals and closely approximate HRRP with high probability. This is a convex problem that conveniently reduces to a linear program known as Basis Pursuit (BP) \cite{9}; ii) greedy algorithms. one representation of those is orthogonal match pursuit (OMP) \cite{14}-\cite{15}; Considering it has a substantial gap between the computational cost of OMP and the cost of BP, we develop faster reconstruct algorithms than OMP algorithm in the paper. For describing convenience, the measurement matrix in measurement system (see Eq. (3)) is called dictionary. Each column in dictionary is called an atom. Meanwhile, it calls that it is $K$-sparse if containing $K$ nonzero entries in a vector.\par
However, in the GTD model with multiple scatterers, a few scattering mechanisms should be considered. With the increase of atom number in dictionary, the computational cost increases. Although a simplified scattering model can be used to approximate multiple scatterers model as discussed in section 3, model mismatch can degrade the success recover probability, which deteriorates the cumulative distribute error (CDE) of SRP. Similar to the SD in \cite{16}-\cite{17} which is used to mitigate inter-atom interference (IAI), the SD is introduced to mitigate model mismatch in this paper. To the authors' knowledge, in the previous work, there is not report in GDT model yet. Using SD, it can reduce computational complexity and mitigate the model mismatch so as to improve the recover probability of SRP. The main contributions of the paper are three aspects. Firstly, for measurement data loss, it adapts sparse property of HRRP to synthesize range profile. In the second, it mitigates model mismatch by introducing SD. Thirdly, an improved fast algorithm (i.e., OMP-SD) is proposed.\par
The paper is organized as follows. In section 2, it first presents the GTD scattered model in frequency domain and then, establishes measurement system in stepped frequency radar (SFR). After that, it briefly reviews existing algorithms to solve the model and presents approximate OMP algorithm (A-OMP) in section 3. In section 4, it presents a strategy to construct SD. It mitigates model mismatch effectively. Besides, a fast algorithm (OMP-SD) to synthesize HRRP is proposed. Monte Carlo simulations illustrate the performance of the proposed algorithm both in AWGN and noiseless situation respectively, in section 5. Finally, some conclusions and further work are provided in Section 6.

Notation: It denotes vectors and matrices by boldface lowercase and uppercase letters, respectively. Uppercase Greek letters also represent matrix in this paper. $(\cdot)^T$ denotes the transpose operation, $(\cdot)^H$ denotes the conjugate transpose operation, Further, $\Vert \cdot \Vert_2$ refers to the $l_2$ norm for vectors. $\Vert \cdot \Vert_{\infty}$ refers to the $l_\infty$ norm for vectors. The $vec(\cdot)$ operator vectorizes a matrix by stacking its columns.  $\mathbf{R} \in \mathbb{R}^{L \times M} $ and $ \mathbf{R} \in \mathbb{C}^{L \times M}$ denote a real-valued and complex-valued matrix and let $\Re\{ \cdot \}$  and $\Im\{ \cdot \}$ be real part and imaginary, respectively. $(\cdot)^+$ denotes the M-P generalized inverse.\par
\textbf{2. ~Problem Formulation}\par
In this section, it briefly presents the GTD scatter model of SFR return signal. SF pulse trains are created by transmitting a train of $M$ identical baseband pulse with different carrier frequencies. The carrier frequency of the $m$-th $(m=0,1,\cdots,M-1)$ pulse is $f_m=f_0+ \Delta f$, where $f_0$ is the initial frequency and $\Delta f$ is the frequency step size. In the stretch processing \cite{20,21}, the range resolution is $\Delta r = c/(2M\Delta f)$, and the ambiguous range $\Delta R = c/(2\Delta f)$ ($c$ is the speed of light). The two-dimensional geometry of the radar scenario is shown in Fig. 1. For the convenience of signal modeling and derivation, it is assumed that the target is stationary and it falls in the range gate $[L,L+L_0]$ in one CPI, where, $L = Q \Delta R$, and $L_0=N\Delta r$). ($Q$ and $N$ are nonnegative integers). Meanwhile, it assumes that the target can be present only the grid points and let us discretize the range space by $\Delta r$ in $L_0$.

\begin{figure}[!t]
\centering
\includegraphics[width=3.5in]{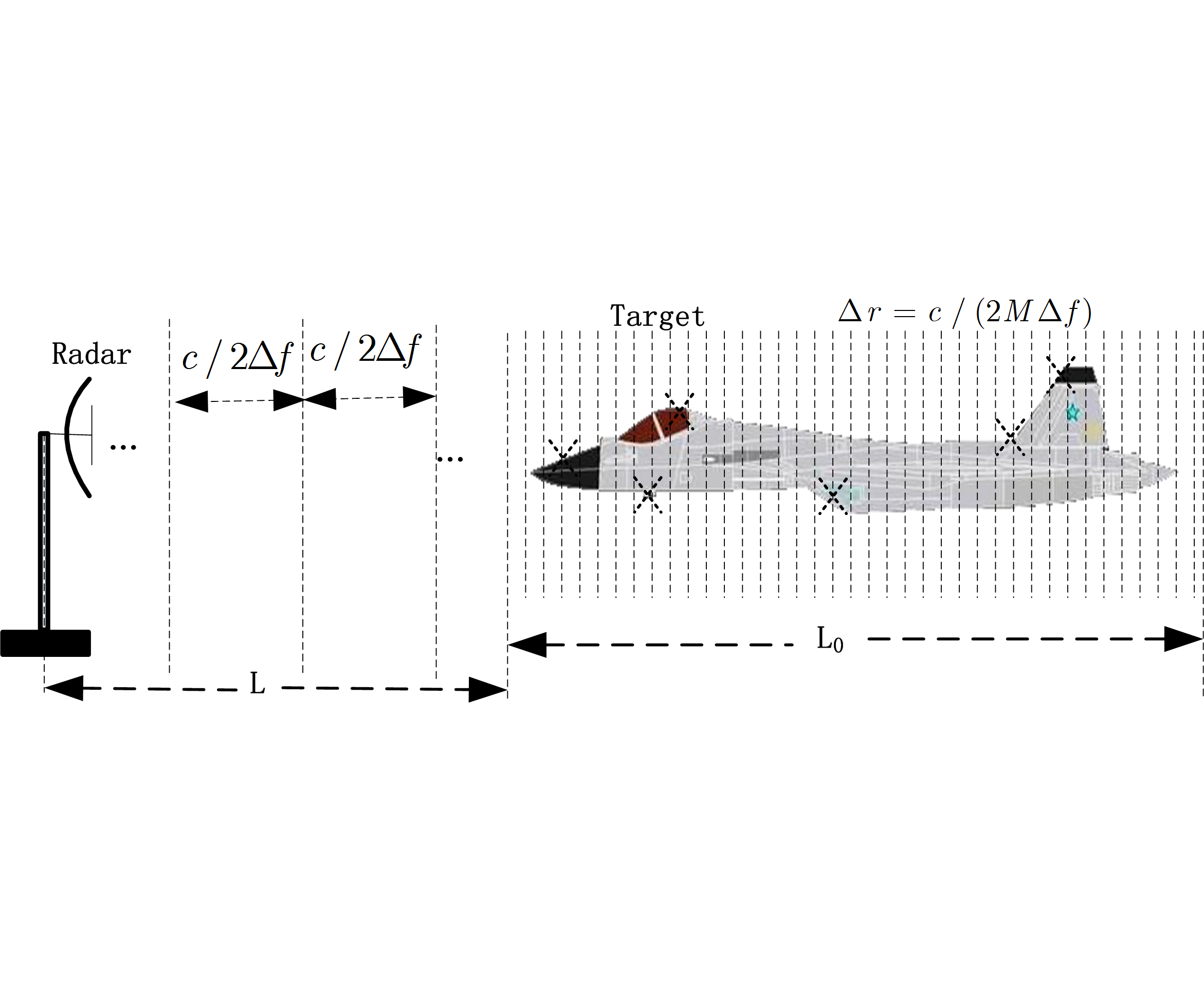}\vspace{-20.0mm}
\caption{Scenario of a target in SFR}
\label{fig_sim} \vspace{-3.0mm}
\end{figure}

In one aspect angle, the parametric GTD scatter model of SFR at frequency $f_m$ can be represented as follows \cite{18}-\cite{19}, \cite{22},
\begin{equation}
y_m =
 \sum_{d=1}^D \sum_{l=1}^NG_d \left(j \frac{f_m}{f_0} \right)^{\alpha_d} \cdot \exp \left\lbrace -j \frac{4\pi}{c} f_m r_{dl} \right\rbrace \cdot x_{dl} + u_m
\end{equation}

where,
\begin{subnumcases}{x_{dl}=}
{1, if ~ scatter~is ~ present~ in ~ r_{dl}}\\
0,otherwise
\end{subnumcases}

In (1), $G_d$, $\alpha_d$ are the complex amplitudes, scattering mechanism of $d$-th scatterer, respectively. $r_{dl}$ denotes range space w.r.t. $l$-th range resolution cell in $m$-th pulse. $D$ is the number of scatterers. $u_m$ is the AWGN with mean zero and variance $\sigma^2$.

The $y_m$ in (1) consists of $N$ uniformly sampled time-domain data from the baseband echo signal of $m$-th pulse $(N=2L_0/(c\Delta t)$ and $\Delta t = 1/(M\Delta f))$. The model can be written into a matrix form as follows,
\begin{equation}
\mathbf{y}=\Phi \mathbf{x} +\mathbf{u}
\end{equation}

  where, $\mathbf{y} \in \mathbb{C}^{M \times 1}$, $\Phi \in \mathbb{C}^{M \times DN}$ and $\mathbf{x} \in \mathbb{R}^{DN \times 1}$ are measurement vector, dictionary and HRRP index of the target, respectively. $\Phi \triangleq [\Phi_{1},\Phi_{2},\cdots,\Phi_{D}]$ and
  \begin{equation}
  [\Phi_d]_{m,n} = G_d[j(1+m\Delta f/f_0)]^{\alpha_d}\cdot \exp(-j2\pi f_m(r_0+n/(M\Delta f)))
  \end{equation}

  $\mathbf{x}=[ \mathbf{x}_{1}^{T} ~ \mathbf{x}_{2}^{T} ~ \cdots ~ \mathbf{x}_D^{T}]^T$, $\mathbf{x}_d \in \mathbb{R}^{N \times 1}$, $\alpha_d \in \Omega$ ($\Omega$ is a set composed of scattering mechanisms), $\mathbf{u} \in \mathbb{C}^{M \times 1}$ is the AWGN vector. For convenience of the later describing, it defines $\Phi \triangleq [\Phi_1,\Phi_2,\cdots,\Phi_D]$ and $\Phi_d \triangleq [\phi_{d1},\phi_{d2},\cdots,\phi_{dN}]$.  $\Phi_{d}$ denotes $d$-th block matrix of $\Phi$. $\phi_{di}$ denotes the $i$-th atom of $\Phi_{d}$. $d \in \Lambda \triangleq \{1,2,\cdots,D\}$. All atoms are normalized throughout the paper. $r_0$ is the radial distance from radar antenna to reference point on the target. In realistic settings, $M<<N<DN$, hence it is an underdetermined system in (3). It is to reconstruct of a high-dimension sparse vector $\mathbf{x}$ from a small number of linear measurements $\mathbf{y}$ and dictionary $\Phi$.\par
\textbf{3. ~The A-OMP Algorithm}\par
To solve an underdetermined system of linear equations in the above form (3), recently $l_1$-norm minimization as an effective technique has attracted attention in the CS community \cite{9}-\cite{13}. It is a convex optimization problem in noise setting:
    \begin{equation}
        (P_1:)
         \min_{\mathbf{x}} \Vert \mathbf{x} \Vert_1 ~~\textit{subject to}~ \Vert \mathbf{y}- \Phi \mathbf{x}\Vert_2 \le \varepsilon
    \end{equation}

For the problem $P_1$, it can be solved by linear programming (LP). Nevertheless, general-purpose LP solvers require about $O(D^3N^3)$ flops. Thus, it is beyond the scope of discussion in the paper. As a matter of fact, many of the applications of $P_1$ can be attacked heuristically by fitting sparse models, using greedy stepwise least squares. A widely used algorithm for sparse signal recovery is the OMP algorithm for the recovery of the support of the $K$-sparse signal in (3), which requires $O(DNMK)$ flops \cite{14}-\cite{15}. For noiseless case, the exactly recovery condition (ERC) of OMP was derived by Troop in 2004. T.cai et al. derived a new ERC both in the bounded noise and Gaussian noise \cite{23}.\par
In (3), considering the computational cost, the sparse solution can be solved with OMP algorithm directly. Rather than minimizing an objective function in (5), OMP constructs a sparse solution to a given problem by iteratively building up an approximation, the vector $\mathbf{y}$ is approximated as a linear combination of a few atoms in dictionary $\Phi$, where the $active set$ of atoms to be used is built column by column, in a greedy fashion. At each iteration, a new atom that best correlates with the current residual is added to the $active set$. The standard OMP algorithm can be found in \cite{14}.\par
For GTD multi-scattering model, a few scatter mechanisms are considered. It increases the atom number in dictionary, and hence it increases computation. To decrease computation caused by multiple scatterers, it's a straight way to synthesize range profile of target that using a single scattering mechanism instead of multiple scattering mechanisms (i.e., to replace $\Phi$ with $\Phi_d$).\par
Just as the description in section 2, the atoms in dictionary (i.e., columns of $\Phi_d$) are normalized so that $\Vert \phi_{di} \Vert_2 =1$, for $ i=1,2,\cdots,N$. It denotes by $c \subseteq  S \triangleq \{1,2,\cdots,N\}$ the support of $\mathbf{x}_d$, which is defined as the set of indices corresponding to the nonzero components of $\mathbf{x}_d$. For matrix $\Phi_d$, $\Phi_d(c)$ denotes the matrix formed by picking the atoms of $\Phi_d$ corresponding to indices in set $c$. Following the same convention as section 2, $\phi_{di}$ represents the $i$-th atom of $\Phi_d$. It calls $\phi_{di}$ a correct atom if the corresponding $\mathbf{x}_{di} \ne 0$ and call $\phi_{di}$ an incorrect atom otherwise. With slight abuse of notation, we use $\Phi_d(c)$ to denote both the subset of atoms of $\Phi_d$ with indices and the corresponding block matrix of $\Phi_d$. A detailed description of approximation orthogonal matching pursuit (A-OMP) algorithm is presented as follows.

\begin{algorithm}
\caption{:A-OMP}
\label{alg:A-OMP}
\begin{algorithmic}[1]       
\REQUIRE ~~\\      

The measurement vector, $\mathbf{y}$; \\
The dictionary, $ \Phi_d$, $d \in \Lambda$; \\
the error threshold, $\epsilon$; \\
\ENSURE ~~\\     

\STATE Initialize the residual $\mathbf{r}_0 = \mathbf{y}$ and initialize the subscript set of selected atom $c_0$ is empty. Set $i=1$.

\STATE Find the atom $\phi_{t_i}$  that solves the maximization problem
\[ t_i \triangleq \max_t \vert \phi_{dt}^H \mathbf{r}_{i-1} \vert , (t \in S, ~\phi_{dt}~ is ~the ~tth ~atom ~in ~ \Phi_d)\]
and update $c_i = c_{i-1} \cup \{t_i\}$.

\STATE Let $\mathbf{P}_i = \Phi_d(c_i)( \Phi_d(c_i)^H \Phi_d(c_i))^{-1} \Phi_d(c_i)^H$.
Denote the projection onto the linear space spanned by the elements of $\Phi_d(c_i)$. Update $\mathbf{r}_i = (\mathbf{I}- \mathbf{P}_i) \mathbf{y}$.

\STATE If the stopping condition is achieved (i.e., $\Vert \mathbf{r}_i \Vert_2 \le \epsilon$), go to 5. Otherwise, set $i= i+1$ and return to 2.
\STATE Pick out the range scattering cells w.r.t set $c_i$.\label{code:fram:extract5}
\STATE Calculate the scattering intensity in these range cells determined in the previous step with $\mathbf{P}_i^+\mathbf{y}$.
\STATE Reconstruct SRP using the scattering intensity and range scattering cells.
\STATE Return SRP.
\end{algorithmic}
\end{algorithm}

Similar to OMP, the A-OMP is a stepwise forward selection algorithm and is easy to implement. A key component of A-OMP is the stopping rule which depends on the noise structure. In the noiseless case the natural stopping rule is $\mathbf{r}_i = 0$. That is, the algorithm stops whenever $\mathbf{r}_i = 0$ is achieved. In this paper, both noiseless and the case of AWGN in which $\mathbf{u}_i \sim \mathit{N}(0,\sigma^2)$ are considered. The stopping rule for each case and the properties of the resulting procedure are discussed in article \cite{14}.\par
As a special case of multi-scattering center, for a single scattering mechanism, the A-OMP algorithm procedure is the same as OMP, but it has a significant different physical meaning. Because the dictionary $\Phi_d$ just as a sub-block of $\Phi$ in (3). Thus, it is called approximate OMP in the paper. Once the subscribe set is determined with A-OMP or OMP, the SRP can be obtained with Least-Square (LS) solution.\par

\textbf{4. ~OMP Algorithm via Sensing Dictionary (OMP-SD)}\par
In (3), the sparse vector $\mathbf{x}$ can be obtained by OMP or A-OMP directly. However, two major problems cannot be avoided in this case. For OMP, it has to search all atoms in dictionary $\Phi$ to find the best matched atom at each iteration (in the paper, the dictionary is $M$-by-$DN$ dimension matrix); For A-OMP, it just needs to find the best matched atom in $M$-by-$N$ dimension dictionary at each iteration, but it leads to model mismatch and increases CDE of SRP. Thus, an improved algorithm via sensing dictionary (i.e., OMP-SD) is developed to overcome drawbacks of both OMP and A-OMP. There are two advantages with OMP-SD to synthesize HRRP. On the one hand, as a result of the atoms in SD are independent on scattering mechanisms, it mitigates model mismatch. On the other hand, it reduces computation because of the searching dimensional of dictionary reduced from $M$-by-$DN$ down to $M$-by-$N$.\par
\textbf{4.1. ~Dictionary Pre-processing}\par
For the convenience of following analysis, the dictionary $\Phi$ of in (3) are divided into $D$ $M$-by-$N$ dimensional block matrix firstly, which are denoted by $\Phi_1,\Phi_2,\cdots,\Phi_D$ and each block matrix $\Phi_d$, $d \in \Lambda$, corresponds to a different scattering mechanism. The Eq. (3) can also be rewritten as,
\begin{equation}
\mathbf{y} = [\Phi_{1} \vert \Phi_{2} \vert \cdots \vert \Phi_D] \mathbf{x} + \mathbf{u}
\end{equation}

In (6), it considers $\Phi_1$ (w.r.t the 1st scattering mechanism) as an example. In ideal condition, the Gram matrix $ \Phi_1^H \Phi_d = \mathbf{I}$, but it is not the case because the dictionary is over complete, so it has to make $ \Phi_1^H \Phi_d \rightarrow \mathbf{I}$, ($d \in \Lambda$ and $d\neq 1$), extremely, which needs to solve the problem $\max_{d\in \Lambda} \Vert \mathbf{I}- \Phi_1^H \Phi_d \Vert_{\infty}$. According to the idea, it should find an $M$-by-$N$ SD $\mathbf{W}$ ( being the same dimensional as block matrix $\Phi_d$ ), which is independent on scattering mechanisms. The SD can be found by solving the problem $P_2$ of the follows,

\begin{subnumcases}{(P_2:)}
\min_{\mathbf{W}} ~b_1 + \gamma b_2 \\
s.t. ~\Vert \mathbf{I}- diag(\mathbf{W}\Phi_d)\Vert_{\infty} \le b_1\\
s.t. ~\Vert \mathbf{\rho} \Vert_{\infty} \le b_2 \\
\mathbf{\rho} = vec \left( (\mathbf{W}^H \Phi_d)_{k,l}\right) , k \ne l \\
d \in \Lambda
\end{subnumcases}

In (7a), $\gamma$ is the regular factor. It sets 0.5 in the paper. Both $b_1$ and $b_2$ are unknown but determined variables. They reflect the IAI level between $\mathbf{W}$ and dictionary $\Phi_d$. As the problem $P_2$ is a convex problem, the sensing dictionary $\mathbf{W}$ can be obtained offline with efficient algorithms. There are many software pockets to solve the problem such as cvx \cite{24} etc.\par
In $P_2$, for the first constraint, it means maximizing correlation of diagonal elements of matrix which is conjugate and transpose operation of SD multiplied by $\Phi_d$. For the second constraint, it means minimizing correlation of off-diagonal elements of matrix which is conjugate transpose of SD multiplied by $\Phi_d$. Done with it like this, the sensing dictionary $\mathbf{W}$ is insensitive to scattering mechanisms. In other words, the model mismatch can be imitated greatly. In the following, it illustrates problem $P_2$ being a convex problem. Noting that $\mathbf{W}$ and $\Phi_d$ are divided column by column and $\mathbf{W} \triangleq [\mathbf{w_1},\mathbf{w_2},\cdots,\mathbf{w}_N]$, $\Phi_d \triangleq [\phi_{d1},\phi_{d2},\cdots,\phi_{dN}]$.
Denoting column vector $\tilde{\phi}_{dl}$, $\hat{\phi}_{dl}$ and $\tilde{\mathbf{w}}_l$ as follows, respectively,
\begin{equation}
\tilde{\phi}_{dl} = \left[\Re(\phi_{dl})^T  ~~\Im(\phi_{dl})^T \right]^T
\end{equation}

\begin{equation}
\hat{\phi}_{dl} = \left[\Im(\phi_{dl})^T  ~~-\Re(\phi_{dl})^T \right]^T
\end{equation}

\begin{equation}
\tilde{\mathbf{w}}_l = \left[\Re(\mathbf{w}_l)^T  ~~\Im(\mathbf{w}_l)^T \right]^T
\end{equation}

let,
\begin{equation}
f_1(\mathbf{W}) = \left\Vert \mathbf{I} - diag(\mathbf{W}^H \Phi_d) \right\Vert_{\infty} - b_1
\end{equation}

Noting that the first inequality constrains in problem $P_2$,
\begin{equation}
\left\Vert \mathbf{I} - diag(\mathbf{W}^H \Phi_d) \right\Vert_{\infty} \le b_1
\end{equation}

which is equivalent as the following constraints,
\begin{equation}
\left\Vert \substack { 1- \mathbf{w}_1^H \phi_{d1}\\
 1- \mathbf{w}_2^H \phi_{d2}\\
 \vdots \\
1- \mathbf{w}_N^H \phi_{dN}} \right\Vert_\infty \le b
  \end{equation}

Without loss of generality, it is supposed that the absolute of $l$th component is the largest in (12), so it has,
\begin{equation}
\left\Vert \substack { 1- \mathbf{w}_1^H \phi_{d1}\\
 1- \mathbf{w}_2^H \phi_{d2}\\
 \vdots \\
1- \mathbf{w}_N^H \phi_{dN}} \right\Vert_\infty = \vert  1-\mathbf{w}_l^H \phi_{dl}\vert
  \end{equation}

Using equation (9), (10) and (11), we can get
\begin{equation}
\left\vert \mathbf{w}_l^H \phi_{dl} \right\vert = \left\vert \tilde{\mathbf{w}}_l \tilde{\phi}_{dl} + j \cdot \tilde{\mathbf{w}}_l \hat{\phi}_{dl} \right\vert
= \left\Vert \tilde{\mathbf{w}_l} ~[\tilde{\phi}_{dl} ~~\hat{\phi}_{dl}] \right\Vert_2
\end{equation}

Obviously, it is a second order cone about $\mathbf{W}$. Hence, $f_1(\mathbf{W})$  is a convex function about $\mathbf{W}$.
For the second constraint condition, let
\begin{equation}
  f_2(W) = \left\Vert  vec \left( (\mathbf{W}^H \Phi_d)_{k,l} \right) \right\Vert_{\infty} - b_2 ~~(k \ne l)
  \end{equation}

Obviously,
\begin{equation}
\left\Vert  vec \left( (\mathbf{W}^H \Phi_d)_{k,l} \right) \right\Vert_{\infty} = \max_{k,l} (\mathbf{W}^H \Phi_d)_{k,l} =\vert \mathbf{w}_k^H \phi_{dl}\vert
  \end{equation}

Done with similar derivation procedure of $f_1(\mathbf{W})$, it is easy to show that $f_2(\mathbf{W})$ is also a convex function about $\mathbf{W}$ and objective function is affine function w.r.t. $b_1$, $b_2$ for a given $\gamma$. Hence $P_2$ is convex.\par

\textbf{4.2. ~The Proposed Algorithm}\par
The proposed algorithm (OMP-SD) is also a greedy algorithm but different from OMP and A-OMP. For OMP-SD, at each iteration, it requires a two-step search to select an atom. First, it determines the offset index of atom in SD $\mathbf{W}$, which is not sensitive to scatter mechanism. And then, it further to determine the specific scatter mechanism in dictionary $\Phi$. After the two-step procedures, an actual atom is picked out. The OMP-SD is described as follows.

\begin{algorithm}
\caption{:OMP-SD}
\label{alg:A-OMP}
\begin{algorithmic}[1]       
\REQUIRE ~~\\      
The measurement vector, $\mathbf{y}$; \\
The dictionary, $ \Phi_1, \Phi_2 \cdots,\Phi_D$, $\mathbf{W}$ \\
the err threshold, $\epsilon$; \\
\ENSURE ~~\\     
\STATE To initialize the residual $\mathbf{r}_0=\mathbf{y}$ and initialize the subscribe set $c_0$ is empty. set $i=1$.
\STATE To find the matrix $\Gamma_{t_i}$ that solves the maximization problem
\[ t_i \triangleq \max_t \left\vert \mathbf{w}_t^H \mathbf{r}_{i-1} \right\vert \]
where, \[ \Gamma =\left[ \Phi_{1}(t_i) ~~\Phi_{2}(t_i) ~~\cdots ~~\Phi_D(t_i)\right] \]
\STATE To solve the maximization problem
\[ \xi = \max_{d} \left\vert \Gamma^H \mathbf{r}_{i-1} \right\vert , d \in \Lambda \]
and update $c_i = c_{i-1} \cup \{ \xi_i\}$. Where $\xi_i = (\xi -1)N + t_i$.
\STATE Let $\mathbf{P}_i = \Phi(c_i)( \Phi(c_i)^H \Phi(c_i))^{-1} \Phi(c_i)^H$
denote the projection onto the linear space spanned by the elements of $\Phi(c_i)$. Update $\mathbf{r}_i = (\mathbf{I}-\mathbf{P}_i)\mathbf{y}$.
\STATE If the stopping condition is achieved (i.e., $\Vert \mathbf{r}_i \Vert_2 \le \epsilon$), go to 6. Otherwise, set $i= i+1$ and go back to 2.
\STATE Pick out the range scattering cells w.r.t set $c_i$.\label{code:fram:extract5}
\STATE Calculate the scattering intensity in these range cells determined in the previous step with $\mathbf{P}_i^+\mathbf{y}$.
\STATE Reconstruct SRP using the scattering intensity and range scattering cells.
\STATE Return SRP.
\end{algorithmic}
\end{algorithm}

As far as computational complexity is concerned, it requires $DN$ times correlation operators to select an atom in OMP whileas  it just requires $N+D$ times for the proposed algorithm. So it requires about $O(M(N+D)K)$ flops. It is approximate to the simplified model in which requires $N$ times. Similarly to in section 3 discussed, once the subscribe set is determined with the proposed algorithm, the SRP can be recovered with LS solution, too.\par

\textbf{5. ~Simulation and Experimental Results}\par
In this section, 10000 trails Monte Carlo simulation has been done to illustrate the previous discussions. Assume the SFR operates at the following condition. Five scattering mechanisms are considered (i.e., $\alpha_d \in \Omega \triangleq \{-1,-0.5, 0, 0.5, 1\}$). An example scattering geometrics and the corresponding scattering parameters are shown in Tab.~\ref{tab:table_example}.

\begin{table}
\centering
\caption{Geometry parameters for example scattering geometries \cite{17}}
\begin{tabular}{|c|c|}
\hline
\textrm{Value of} $\alpha_d$ & \textrm{Scatter mechanisms} \\ \hline
-1   & corner diffraction\\ \hline
-0.5 & edge diffraction \\ \hline
0    & point diffraction; straight edge specular \\ \hline
0.5  & singly curved surface reflection\\ \hline
1    & late plate at broadside; dihedral\\ \hline
	\end{tabular}
\label{tab:table_example}
\end{table}

In (4), it is assumes that target is stationary in one CPI, and the distance $L$ in Fig. 1 is regarded as constant. In simulation, set the $r_0=0$ and hence, the $m$th row and $n$th column element in (4) is rewritten as
  \begin{equation}
  [\Phi_p]_{m,n} = G_p[j(1+m\Delta f/f_0)]^{\alpha_p}\cdot \exp(-j2\pi f_m n/(M\Delta f)))
  \end{equation}

The range of the measured frequency band is from L band to S band ( i.e., from 1GHz to 4GH ), where the start frequency is $f_0=1$GHz and frequency step size $\Delta f=10$MHz. The number of pulses $M = 300$. And it assumes that the target is 5m length. Five scatterers are located on 0.3m, 0.85m, 2.0m, 3.25m and 4m to target front-end, respectively. All scatterers have same intensity. What's more, it assumes that the stationary scatterer centers are present on the grid points. In each measurement, only 30 returned pulses are measured in one CPI (i.e., 300 pulses). The measurement vector $\mathbf{y}$ is contaminated by AWGN with $SNR=20$dB, 15dB, 10dB, 5dB and noiseless situation, respectively. In order to explain the essence of model mismatch, mutual incoherence property (MIP) is introduced which is defined as the same as in article \cite{25},
\begin{equation}
\mu (\Phi) \triangleq \max_{\substack{1 \leqslant i,j \leqslant n \\ i \ne j}}
\frac{\left\vert {\phi}_i^H \phi_j \right\vert}{\Vert {\phi}_i^H\Vert_2 \cdot \Vert \phi_j \Vert_2}
\end{equation}

Noting that each atom in dictionary is normalized, hence Eq. (19) can be rewritten as another form,
\begin{equation}
\mu (\Phi) \triangleq \max_{\substack{1 \leqslant i,j \leqslant n \\ i \ne j}}
{\left\vert {\phi}_i^H \phi_j \right\vert}
\end{equation}

\begin{figure}[!t]
\centering
\includegraphics[width=3.5in]{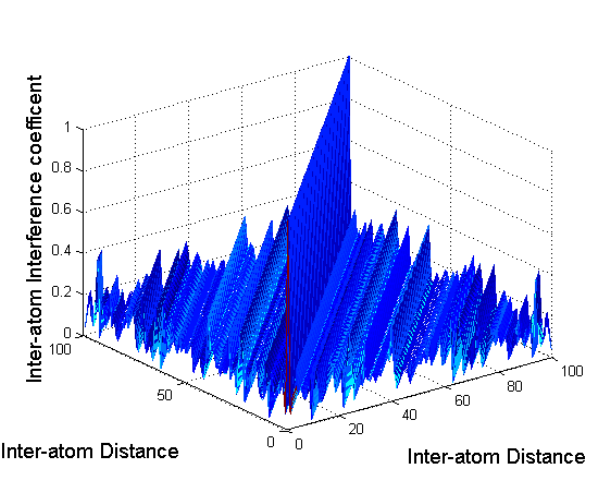}\vspace{0.0mm}
\caption{IAI in original dictionary}
\label{fig_sim} \vspace{0.0mm}
\end{figure}

\begin{figure}[!t]
\centering
\includegraphics[width=3.5in]{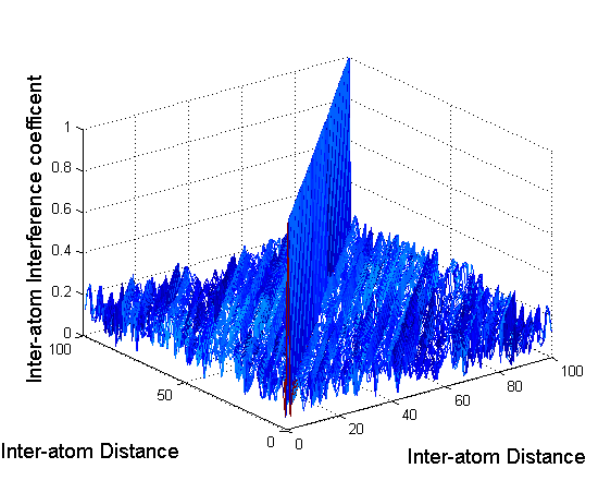}\vspace{0.0mm}
\caption{IAI in sensing dictionary}
\label{fig_sim} \vspace{0.0mm}
\end{figure}

\begin{figure}[!t]
\centering
\includegraphics[width=3.5in]{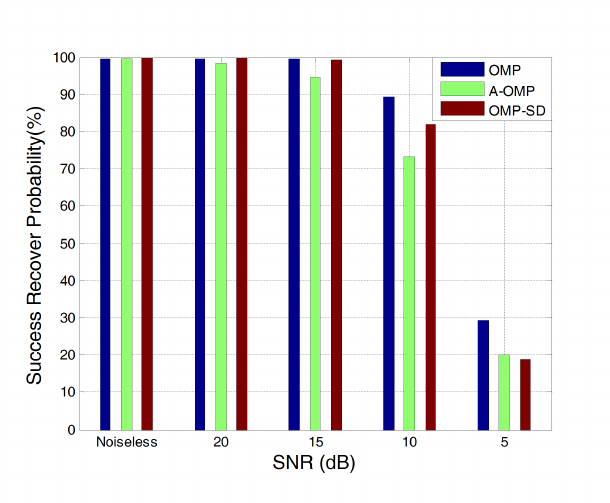}\vspace{0.0mm}
\caption{Success recover probability w.r.t SNR}
\label{fig_sim} \vspace{0.0mm}
\end{figure}

\begin{figure}[!t]
\centering
\includegraphics[width=3.5in]{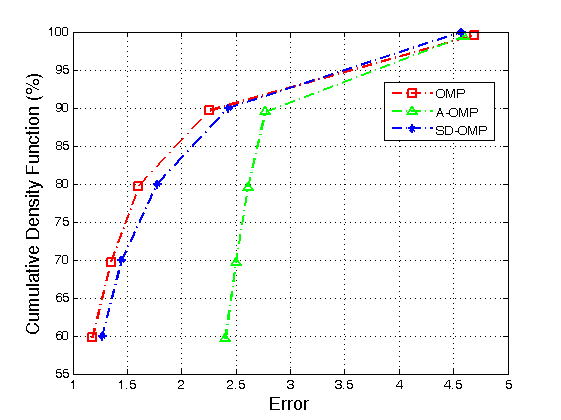}\vspace{0.0mm}
\caption{Cumulative distribute error in noiseless}
\label{fig_sim} \vspace{0.0mm}
\end{figure}

\begin{figure}[!t]
\centering
\includegraphics[width=3.5in]{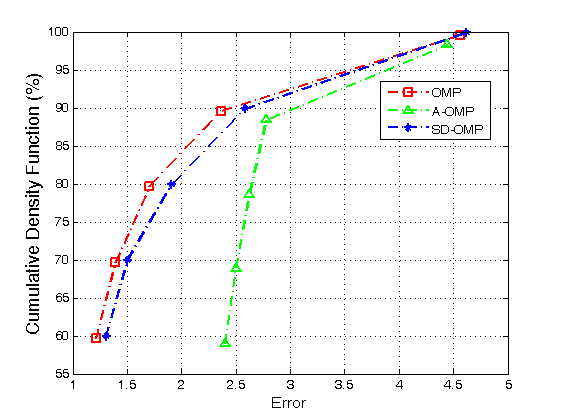}\vspace{0.0mm}
\caption{Cumulative distribute error $SNR=20$dB}
\label{fig_sim} \vspace{0.0mm}
\end{figure}

\begin{figure}[!t]
\centering
\includegraphics[width=3.5in]{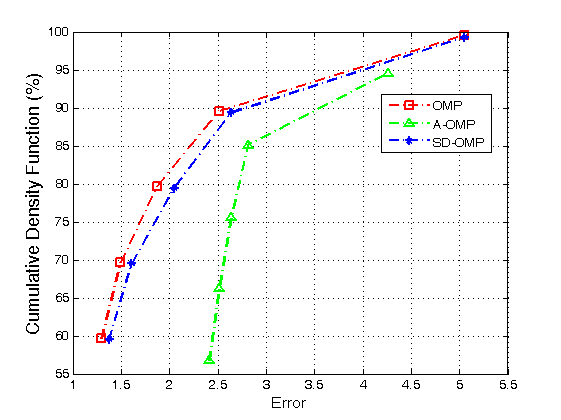}\vspace{0.0mm}
\caption{Cumulative distribute error $SNR=15$dB}
\label{fig_sim} \vspace{0.0mm}
\end{figure}

\begin{figure}[!t]
\centering
\includegraphics[width=3.5in]{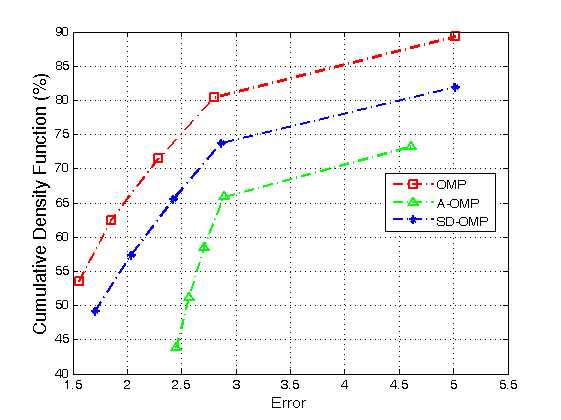}\vspace{0.0mm}
\caption{Cumulative distribute error $SNR=10$dB}
\label{fig_sim} \vspace{0.0mm}
\end{figure}

\begin{table}
\centering
\caption{Time consuming simulation results}
\begin{tabular}{|c|c|}
\hline
\textrm{SRP Algorithm} & \textrm{time(s)} \\ \hline
OMP   & 1044\\ \hline
A-OMP & 109 \\ \hline
OMP-SD    &121 \\ \hline
\end{tabular}
\label{tab:table_example}
\end{table}

Simulation results are shown in Fig. 2 $\sim$ Fig. 8. and it presents several remarks in the following.

\textbf{Remark 1:} The IAI between the original dictionary and SD are shown in Fig. 2 and Fig. 3, respectively. In the original dictionary, the IAI minimum is 0.0762 and the maximum of inner-atom cross-correlation is 0.3872. However, both of them are 0.2248 (i.e., $b_1 = b_2 = 0.2248$ in (7a)) for the sensing dictionary. It improves the autocorrelation between atoms and mitigates model mismatch at the same time.

\textbf{Remark 2:} Fig. 4 shows that the success recovery probability is a monotonic decreasing relative to SNR for the three algorithms (i.e., OMP, A-OMP and OMP-SD). It is easy to understand that the OMP has the best recovery performance because it is match model and the A-OMP has worst recovery performance because of its model mismatch. However, the proposed method (OMP-SD) has an approximate performance compared to OMP and approximate computational complexity to A-OMP and it is confirmed in Tab.~\ref{tab:table_example}. It should be noted that there is an exception for small SNR ($<$ 10dB). When SNR is 5dB in Fig. 4, all of the three algorithms have a lower success recovery probability (less than 30 percent). Hence, objectively speaking, it is a drawback for these algorithm. But in moderately high SNR settings (i.e., greater than 15dB), the proposed algorithm has outstanding performance.

\textbf{Remark 3:} For the noiseless and three different SNR settings, Fig. 5 $\sim$ Fig. 8 show the cumulative distribute errors (i.e., CDE). It is widely used to evaluate recover performance in CS community such as \cite{25}. From Fig. 5 $\sim$ Fig. 8 we can see the match model is best, while the mismatch model is worst although it requires least computation amount. However, OMP-SD shows that it has an approximate values of CDE compared to OMP. However, it has to point out that all of the three algorithms are not suitable for low SNR ($<$ 10dB) settings.

\textbf{Remark 4:} In Tab.~\ref{tab:table_example}, for the same computer platform, there are sums of 10000 Monte Carlo trails time consuming results for OMP, A-OMP and OMP-SD, respectively. It confirms that OMP has the most computational cost but it is approximately computational cost between A-OMP and OMP-SD. Both of them have much lower computation cost compared with OMP.\par

\textbf{6. ~Conclusion and Future Work}\par
In this paper, a fast algorithm to synthesize range profile is proposed. For the SFR system in GTD model, the HRRP synthesis can be converted to solve a sparse approximation problem over redundant dictionaries. Different from A-OMP, the model mismatch is mitigated with SD. Better than OMP, the computational complexity is reduced. Finally, simulation results show the proposed algorithm is valid for both noiseless and noisy settings.\par
However, it just presents the fast algorithm to recover parameters in GTD model. In the future work, it will derive the ERC. Secondly, in this paper, it just presents how to estimate the support of sparse vector. For the scatterer intensity in GTD model, it simply exploits a plain LS method to recover it. In fact, there are a great many algorithms to estimate its magnitude such as biased estimation techniques \cite{27}, etc. All of them will be considered in our future work.\par

\section*{Acknowledgment}

The authors would like to thank the anonymous reviews for their comments that help to improve the quality of the paper.
This research was supported by the National Natural Science Foundation of China (NSFC) under Grant 61172140, and '985' key projects
for excellent teaching team supporting (postgraduate) under Grant A1098522-02. Yipeng Liu is supported by FWO PhD/postdoc grant: G.0108.11 (Compressed Sensing).

\ifCLASSOPTIONcaptionsoff
  \newpage
\fi

\end{document}